**Next-generation MRD assays: do we have the tools to evaluate them properly?**

**Dan Stetson, MS[1]; Paul Labrousse, MS[1]; Hugh Russell, PhD[1]; David Shera, ScD[2]; Chris Abbosh, MD, PhD[3]; Brian Dougherty, PhD[1]; J. Carl Barrett, PhD[1]\*; Darren Hodgson, PhD[3]; James Hadfield, PhD[4]**

[1]Translational Medicine, Oncology R&D, AstraZeneca, Waltham, MA, USA; [2]Oncology Biometrics, AstraZeneca, Gaithersburg, MD, USA; [3]Cancer Biomarker Development, Oncology R&D, AstraZeneca, Cambridge, UK; [4]Translational Medicine, Oncology R&D, AstraZeneca, Cambridge, UK.

\*Retired from AstraZeneca.

**Corresponding author:**

James Hadfield, PhD, Translational Medicine, Early Oncology, Oncology R&D, AstraZeneca, Cambridge, UK, 07795580274, james.hadfield@astrazeneca.com, Twitter=@coregenomics.

**ctDNA detection of molecular residual disease (MRD) in solid tumors: a new clinical standard**

Circulating tumor DNA (ctDNA) is a component of cell-free DNA (cfDNA) that originates from tumor tissue in the plasma of cancer patients.[1, 2] The majority of DNA found in the bloodstream is derived from normal cells, but a small percentage originates from tumor cells in patients with early-stage cancer.[3] Although ctDNA is a broadly applicable tumor biomarker, the absolute amount of ctDNA varies significantly across indication and stage,[4] with the level of ctDNA present in the plasma being partly associated with tumor volume.[5] Because of this, current ctDNA genotyping assays can lack clinical sensitivity due to low ctDNA shedding, particularly in patients with early-stage disease or in the post-therapy setting.[6, 7]

The analysis of ctDNA with next generation sequencing (NGS) technologies has rapidly developed as an important non-invasive oncology diagnostic tool.[8-10] Much of the research and/or clinical development focus has been on ctDNA analysis as a diagnostic test for which assays such as Guardant360 (Guardant Health),[11] FoundationOne Liquid CDx (Foundation Medicine, Inc,),[12] the AVENIO ctDNA liquid biopsy platform (Roche Sequencing

Solutions),[13] tagged-amplicon deep sequencing (TAm-Seq),[9] and duplex sequencing (Duplex-Seq)[14] have been developed for sensitive mutational analysis of cancer genes in patients with advanced-stage disease. However, most of the assays have fixed content focused on prevalent somatic genome variations (single nucleotide variants [SNV], copy number variants [CNV], and structural variants [SV]), whereas ctDNA contains multiple biologic features beyond these somatic variants, such as indication-specific DNA methylation signatures,[15-18] fragment size distribution,[19] and nucleosome positioning patterns (fragmentome)[20-22] and histone modifications,[23] all of which have the potential to increase both the sensitivity and specificity of ctDNA analysis.

The detection of ctDNA in patients with early-stage solid tumors who have completed curative-intent therapy associates strongly with patient outcomes, and emerging data on clinical utility in tumor types such as colorectal cancer[24, 25] and muscle invasive bladder cancer[26] suggest these technologies might be integrated into routine clinical decision-making in the near future. The clinical benefit of ctDNA analysis in patients with advanced-stage disease has been demonstrated in numerous studies.[8, 10, 27] However, the assays used in the advanced-stage setting lack the sensitivity required to detect ctDNA in the early-stage or post-curative-intent-therapy settings, and this unmet clinical need is driving the adoption of novel sensitive assays for the detection of MRD in these settings.[28]

**Increased assay sensitivity**

MRD detection requires sufficient sensitivity to capture low levels of ctDNA in order to enable detection of clinically occult metastatic disease in advance of routine surveillance imaging. The minimum threshold is 0.01% cTAF (circulating Tumor Allele Fraction) for post-surgical MRD based on initial clinical validity data for MRD detection in solid tumors arising from assays exhibiting a 95% limit of detection of approximately 0.01%.[7, 25, 29] However, early assays were limited in their detection sensitivity because of the relatively small number of variants assessed and/or limited capability to control background artefactual errors.[5, 6, 30] Multiple vendors have developed MRD NGS assays that offer improved sensitivity, primarily by tracking increasing numbers of mutations – from tens (Archer/Invitae, Natera, Inivata/NeoGenomics) to thousands (C2i Genomics, Personalis) of variants – identified from whole-exome or whole-genome sequencing of formalin-fixed paraffin-embedded (FFPE) tumor-tissue and tracked in plasma using personalized ctDNA amplicon or capture panels, and by leveraging novel approaches to control background error rates such as duplex sequencing[31, 32 Schmitt 2012, Bae 2023] and targeting of low-noise events such as phased variants.[33] These tumor-informed and personalized MRD assays have greater sensitivity than fixed-content genotyping assays, but the requirement for tumor sequence information makes them challenging to deliver in a time-sensitive and cost-effective manner. C2i's bioinformatic assay personalization removes the need for a bespoke wet-lab panel, but the assays are still tumor-informed and focused on SNV, insertion/deletion variants (InDel), and SV genomic variants.[34]

Most recently, assays using other biologic features of ctDNA have been developed at the limit of detection (LOD) required for clinical utility. This latest generation of assays, including GuardantINFINITY (Guardant Health),[35] the GRAIL multicancer early detection (MCED) diagnostic test,[36] and the DELFI (DNA evaluation of fragments for early interception) approach,[20] incorporate additional features such as methylation signatures and DNA fragment size distributions (fragmentome) to enable tumor-naive and non-personalized MRD evaluation. These assays should be easier to implement than the tumor-informed and personalized assays but may not have the same level of sensitivity.

What level of sensitivity is required is still unclear. Recently, Parsons et al[37] showed that more than half of patients assessed with their MAESTRO assay, an ultrasensitive MRD method, had tumor fractions below the threshold of most commercially available MRD tests, highlighting the need for more sensitive tests to reduce false MRD-negative calls and to maximize the opportunity to escalate therapeutic intervention in patients with a higher risk of recurrence.

**The challenge of benchmarking new ctDNA MRD assay technologies**

However, rapid development in MRD assay technology results in significant challenges in benchmarking these new technologies for use in clinical trials. Benchmarking ctDNA assays is required to allow selection of the most appropriate method for a clinical study and can assess multiple factors such as LOD, limit of quantification (LOQ), precision, accuracy, and reproducibility.[38] The ctDNA research community, including the US Foundation for the National Institutes of Health (FNIH) Biomarkers Consortium,[39] the Sequencing and Quality Control 2 project,[40] the Friends of Cancer Research ctDNA for Monitoring Treatment Response (ctMoniTR) project,[27] and most recently the EU Innovative Health Initiative GUIDE.MRD project,[41] have put forth, or are evaluating, several types of contrived reference materials including fragmented cell lines and oligonucleotides.[38, 42, 43] The use of these samples to assess ctDNA assay performance has worked well for assays examining genomic features (i.e., somatic variants), but these samples do not carry the disease-specific methylation, fragmentome, or histone-modification features assessed by newer technologies, and they are not commutable to clinical samples (Figure 1A). However, clinical trial samples lack the required volume to perform multiple tests and are not usually consented for assay development.

*'Plasma-in-plasma' reference standards – the 'least worst' option?*

To assess both current and future MRD technologies for use in AstraZeneca clinical trials, we have focused on the development of 'plasma-in-plasma' reference standards, primarily to validate assay LOD. These contrived samples are produced by mixing plasma from cancer patients into high-volume healthy donor plasma, such that they contain all the ctDNA biologic features of clinical samples but in a dilution series of known contrived tumor fraction (cTF). Commercially acquired samples are extensively characterized before reference standard production. Cancer patient samples undergo whole-exome sequencing of FFPE tissue and panel NGS[44] of ctDNA to identify those with high disease burden in plasma using mean variant allele frequency (VAF) of tissue:plasma concordant somatic variants. Gender- and haploid genome equivalent-matched healthy donor plasma, collected as 100 ml serial blood draws to enable the use of single donor plasma as the diluent, is screened for both clonal hematopoiesis of indeterminate potential (by testing the matched PBMCs) and incidental findings. Cancer patient plasma is then diluted into healthy donor plasma to create a 1% cTF sample, confirmed by ddPCR, that is then serially diluted to the levels required. The serial dilution range can be adjusted based on disease indication or any other factors; dilution replicates are used to assess reproducibility and account for stochastic effects from such dilute samples (Figure 1B). Cancer-negative samples (i.e., healthy donor plasma) are also included to minimally evaluate assay specificity.

From our perspective, these 'plasma-in-plasma' reference standards are the least-worst option to assess the broadening spectrum of MRD (and molecular response) assays available. The BLOODPAC consortium has recently developed best practice guidelines for benchmarking mutation analysis of ctDNA[45] in which they describe the need to supplant clinical samples with contrived samples that have high commutability (i.e. functionally comparable to clinical samples). They suggest the use of commercially available contrived reference materials; however, other options include ctDNA from clinical samples (usually impossible to acquire in sufficient volume), fragmented cell line genomic DNA, or cfDNA from cell culture media. These reference materials, which are a valuable resource in the validation of genomic assays, each lack some tumor-specific epigenomic features which reduces their commutability to clinical samples. Whilst genomic variant and DNA size profiles can be almost totally comparable to plasma cfDNA, the epigenomic profile (e.g., DNA methylation, fragmentome, or histone code; Figure 1) is likely to be significantly different and may not be commutable in newer epigenomic MRD assays.

We have successfully used these contrived samples to assess many of the currently available tumor-informed and tumor-naïve MRD assays.[46, 47] Importantly, we have confirmed that these contrived samples are highly commutable to clinical trial samples. The FFPE tumor and normal tissue provide sequencing results similar to the success rates of clinical trial samples and the cfDNA yield is within the range of clinical trial samples reported from both internal and public data.

**Conclusions and future directions**

We believe that the 'plasma-in-plasma' reference standards approach suggested here gets as close as possible to actual clinical samples and results in a highly commutable reference material. We have shown that this novel set of contrived samples can be successfully analyzed by a variety of both tumor-informed and tumor-naive MRD assays. While much remains to be determined for optimal reference materials of MRD methods, we are continuing to improve the generation of these samples to test future assays. Future improvements include the incorporation of more replicates at certain dilution levels to increase our ability to understand the quantitative accuracy of an assay. We hope that sharing our experience and providing opinions from 3 years of MRD technology benchmarking in AstraZeneca Translational Medicine will help guide the community conversation.

**Acknowledgments**

Medical editing support for the development of this manuscript, under the direction of the authors, was provided by Steve Hill, PhD, and James Holland, PhD, of Ashfield MedComms (Macclesfield, UK), an Ashfield Health company, and was funded by AstraZeneca.

**Conflict of interest disclosures:**

D. Stetson, P. Labrousse, H. Russell, D. Shera, C. Abbosh, B. Dougherty, J. C. Barrett, D. Hodgson, and J. Hadfield report current or former employment by, and ownership of stock/shares in, AstraZeneca.

**Figure 1. A. Comparison of the various DNA features across different reference standards. B. Evolution of the contrived samples, showing increased replicates at lower cTF and indication-specific sample sets.** Multiple MATRIX studies have been carried out, each building on the experiences gained in the one before and/or changing to reflect the specific question or indication being assessed.

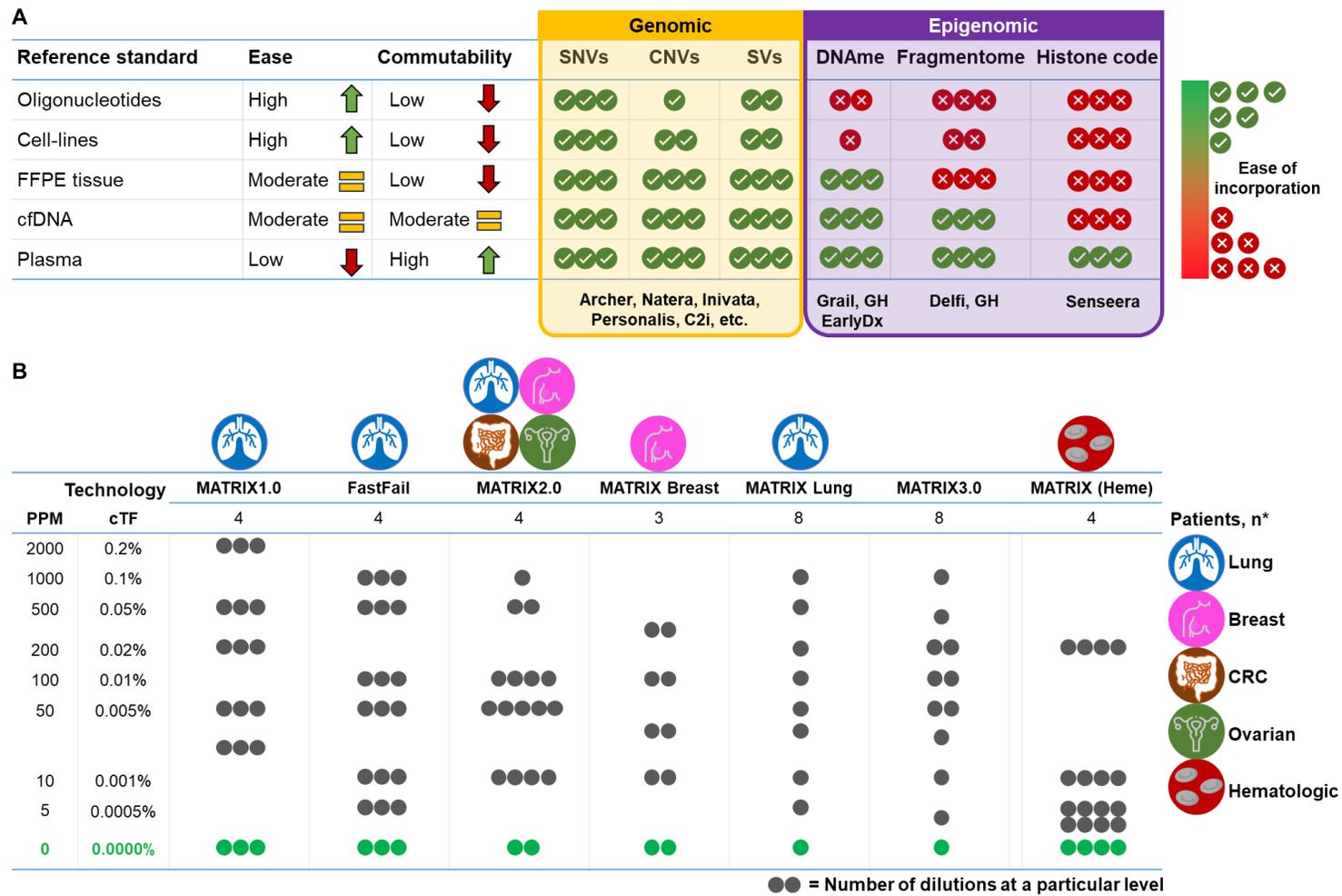

*Number of unique commercial cancer patient samples used. cTF, contrived tumor fraction; PPM, parts per million.